\documentstyle[12pt]{article}

\begin{document}

\setcounter{page}{1}

\pagestyle{plain} \vspace{1cm}
\begin{center}
\Large{\bf Modified GBIG Scenario as an Alternative for Dark Energy}\\
\small \vspace{1cm} {\bf Kourosh Nozari $^{*}$}\quad and \quad {\bf Narges Rashidi}\\
\vspace{0.5cm} {\it Department of Physics, Faculty of Basic
Sciences,\\
University of Mazandaran,\\
P. O. Box 47416-95447, Babolsar, IRAN\\
$^{*}$knozari@umz.ac.ir}

\end{center}
\vspace{1.5cm}

\begin{abstract}
We construct a DGP-inspired braneworld model where induced gravity
on the brane is modified in the spirit of $f(R)$ gravity and stringy
effects are taken into account by incorporation of the Gauss-Bonnet
term in the bulk action. We explore cosmological dynamics of this
model and we show that this scenario is a successful alternative for
dark energy proposal. Interestingly, it realizes the phantom-like
behavior without introduction of any phantom field on the brane and
the effective equation of state parameter crosses the cosmological
constant line naturally in the same way as observational data
suggest.
\end{abstract}
\newpage

\section{Introduction}
According to recent cosmological observations, our universe is
undergoing an accelerating phase of expansion and transition to the
accelerated phase has been occurred in the recent cosmological past
[1]. In order to explain this remarkable behavior and despite the
intuition that this can be achieved essentially only through a
fundamental theory of nature, we can still propose some theoretical
approaches to describe this astonishing feature. It is thought that
the present acceleration of the universe expansion might be due to a
dark energy component in the universe [2]. The simplest way to
describe the accelerated universe is to use the cosmological
constant in the Einstein field equations. However, huge amount of
fine-tuning required for its magnitude and other theoretical
problems such as unknown origin and lake of dynamics make it
unfavorable for this purpose. It is then tempting to find
alternative models of dark energy which have dynamical nature unlike
the cosmological constant. In this respect, it is a challenging job
in theoretical physics to formulate a theory for cosmological
evolution with capability to address the origin of dark energy too.
It is known from cosmological observations that the dark energy
content of the universe has a major contribution on the total energy
budget of the universe. Nevertheless, the usual fields available in
the standard model of particle physics are not sufficient to account
for huge dark energy reservoir in the universe. In this respect, the
concept of dark energy in the Einstein's gravity just with standard
matter or fields cannot be implemented. Therefore, a modification to
the Einstein's field equations either in the geometric or in the
matter sector is essential to accommodate the present cosmological
observations. In this viewpoint, Einstein's field equations can be
reformulated as either $G_{\mu\nu}=8\pi G
(T^{(M)}_{\mu\nu}+T^{(Dark)}_{\mu\nu})$\, ( where $T^{(M)}_{\mu\nu}$
and $T^{(Dark)}_{\mu\nu}$ are energy-momentum tensor of ordinary
matter and dark energy respectively), or the geometric part of these
equations acquire a modification to incorporate the "Dark Geometry"
as $G_{\mu\nu}+G^{(Dark)}_{\mu\nu}=8\pi G T^{(M)}_{\mu\nu}$.\, Both
of these approaches have attracted a lot of attention in the recent
years.

In this respect, braneworld scenarios belong to the second category:
they modify essentially the geometric sector of the Einstein's field
equations. Within a braneworld viewpoint, the
Dvali-Gabadadze-Porrati (DGP) setup provides a new mechanism to
explain the late-time acceleration of the universe based on a
modification of gravitational theory in an induced gravity (IG)
perspective [3]. This new mechanism to explain cosmological
late-time acceleration is more interesting than introduction of an
exotic dark energy and it is able to explain the origin of the dark
energy proposal from a purely gravitational viewpoint. Despite this
astonishing feature, the self-accelerating branch of the DGP
scenario is unstable due to existence of ghosts [4]. However, it
seems that extension of the DGP scenario to more generalized
frameworks such as the model presented in this paper have the
potential to overcome the ghost problem due to their wider parameter
spaces [5]. On the other hand, if we accept the pure DGP setup as
our gravitational theory, we still have some problem in order to
explain all histories of the universe and to match in a smooth way
different phases of the universe evolution. Fortunately, it has been
shown that the normal, non-self-accelerating branch of the DGP
scenario has the potential to explain the phantom-like behavior
without introducing any phantom fields on the brane [6]. Recently,
it has been shown that incorporation of the modified gravity with
$f(R)$ term on the normal branch of the DGP setup has the potential
to realize the late-time acceleration [7]. This interesting
achievement opens new windows on the phantom-like prescription in
DGP-inspired scenarios, one of which is presented in this paper.

From another perspective, in a braneworld scenario the radiative
corrections in the bulk lead to higher curvature terms in the
action. At high energies, the Einstein-Hilbert action will acquire
quantum corrections. The Gauss-Bonnet (GB) combination arises as the
leading bulk correction in the case of the heterotic string theory
[8]. This term leads to second-order gravitational field equations
linear in the second derivatives in the bulk metric which is ghost
free [9,10,11], the property of curvature invariant of the
Gauss-Bonnet term. Inclusion of the Gauss-Bonnet term in the bulk
action results in a variety of novel phenomena which certainly
affects the cosmological dynamics of these generalized braneworld
setup, although these corrections are smaller than the usual
Einstein-Hilbert terms [12-15]. In the presence of the Gauss-Bonnet
(GB) term in the bulk action and induced gravity (IG) on the brane
(GBIG), there are different cosmological scenarios, even if there is
no matter in the bulk [16]. In this paper, we generalize the
previous studies to the case that induced gravity on the brane is
modified in the spirit of $f(R)$ gravity. We show that there are
several interesting features which affect certainly the cosmological
dynamics on the brane. Since the Gauss-Bonnet and induced gravity
effects are related to the two extremes of the scenario
(ultra-violet (UV) and infra-red (IR) limits), inclusion of stringy
effects via Gauss-Bonnet term leads to a \emph{finite density big
bang}. This interesting feature has been explained in a fascinating
manner by $T$-duality of string theory [17].

The motivation for incorporation of the modified induced gravity on
the brane lies in the fact that modified gravity emerges as serious
candidate which is able to addressing the definitive answers to
several fundamental questions about dark energy. For example, the
origin of dark energy may be explained by terms that could be
relevant at late times. Also these terms can be considered as the
source of early time inflation [18]. Therefore, modified gravity is
a natural scenario to have a unified theory to explain both the
inflationary paradigm and dark energy problem. The standard
Einstein's gravity may be modified at low curvature by including the
terms that are important precisely at low curvature. The simplest
possibility is to consider a $\frac{1}{R}$ term in the
Einstein's-Hilbert action [19]. It has been suggested that such a
theory may be suitable to derive cosmological models with late
accelerating phase. Although a theory with $\frac{1}{R}$-term in the
Einstein's gravity accounts satisfactorily the present acceleration
of the universe, it is realized that inclusion of such terms in the
Einstein-Hilbert action leads to instabilities [20]. A modified
theory of gravity which contains both positive and negative powers
of the curvature scalar $R$ namely, $f(R)=R+\alpha R^{m}+\beta
\frac{1}{R^{n}}$ where $\alpha$ and $\beta$ represent coupling
constants with arbitrary constants $m$ and $n$ are considered for
exploring cosmological models [18]. It is known that the term
$R^{m}$ dominates and it permits power law inflation if $1<m\leq2$,
in the large curvature limit. Recently a phenomenologically more
reliable form of $f(R)$ has been proposed in Ref. [21] which we will
consider as a suitable ansatz in our study.

With these preliminaries, the motivation of this work is to
construct a braneworld scenario with induced gravity and
Gauss-Bonnet effect where the induced gravity itself is modified in
the spirit of $f(R)$ gravity. Then we examine cosmological dynamics
in this setup and we show that its account for phantom-like behavior
without introducing any phantom field that violates the null energy
condition neither on the brane nor in the bulk. This model naturally
realizes a smooth phantom divide line crossing by its equation of
state parameter and this crossing occurs in the way that is
supported by observations, that is, from above $-1$ to its below. We
show also that the phantom-like behavior occurs in this setup
without violation of the null energy condition at least in some
subspaces of the model parameter space.

\section{The Setup}

The action of our modified GBIG model contains the Gauss-Bonnet term
in the bulk and modified induced gravity term on the brane as
follows
$$ S=\frac{1}{2\kappa_{5}^{2}}\int d^{5}x\sqrt{-g^{(5)}}\bigg\{R^{(5)}-
2\Lambda_{5}+\beta\bigg[R^{(5)2}-4R_{ab}^{(5)}R^{(5)ab}+R_{abcd}^{(5)}R^{(5)abcd}\bigg]\bigg\}$$
\begin{equation}
+\int_{brane} d^{4} x
\sqrt{-g}\bigg[M_{5}^{3}K^{\pm}+\frac{M_{4}^{2}}{2}f(R)-\lambda+L_{m}\bigg]\
,
\end{equation}
where $\beta(\geq 0)$ is the GB coupling constant, $r_{c}(\geq 0)$
is the IG cross-over scale, $\kappa_{5}^{2}$ is the five dimensional
gravitational constant, $\lambda$ is the brane tension and
$\Lambda_{5}$ is the bulk cosmological constant. Also $\overline{K}$
is the trace of the mean extrinsic curvature of the brane defined as
follows
\begin{equation}
\overline{K}_{\mu\nu}=\frac{1}{2}\lim_{\epsilon\rightarrow
+0}\Bigg(\Big[K_{\mu\nu}\Big]_{y=-\epsilon}
+\Big[K_{\mu\nu}\Big]_{y=+\epsilon}\Bigg)
\end{equation}
and its presence guarantees the correct matching conditions across
the brane. We denote the matter field Lagrangian by $L_{m}$ with the
following energy-momentum tensor
\begin{equation}
T_{\mu\nu}=-2\frac{\delta L_{m}}{\delta g^{\mu\nu}}+g_{\mu\nu}L_{m}.
\end{equation}
The field equations resulting from the action (1) are given as
follows
\begin{equation}
M_{5}^{3}\Big(G_{\,\,B}^{A}+\alpha
H_{\,\,B}^{A}\Big)=\Lambda_{5}\delta_{\,\,B}^{A}+\delta_{\,\,\mu}^{A}
\delta_{\,\,B}^{\nu}\tau_{\,\,\nu}^{\mu}\,.
\end{equation}
The corrections to the Einstein Field Equations originating in the
GB term are represented by the Lovelock tensor [22]
\begin{equation}
H_{\,\,B}^{A}=2RR_{\,\,B}^{A}-4R_{\,\,K}^{A}R_{\,\,B}^{K}-4R^{KL}R_{\,\,KBL}^{A}
+2R^{AKLM}R_{\,\,BKLM}-\frac{1}{2}g_{\,\,B}^{A}L_{GB}.
\end{equation}
The energy-momentum tensor localized on the brane is
$$\tau_{\,\,\nu}^{\mu}=-M_{4}^{2}f'(R)G_{\,\,\nu}^{\mu}-\frac{M_{4}^{2}}{2}\Big[R
f'(R)-f(R)+2\frac{\lambda}{M_{4}^{2}}\Big]\delta_{\,\,\nu}^{\mu}+T_{\,\,\nu}^{\mu}$$
\begin{equation}
+M_{4}^{2}\Big[D^{\mu}D_{\nu}f'(R)-\delta_{\,\,\nu}^{\mu}D^{\mu}D_{\mu}f'(R)\Big],
\end{equation}
where a prime marks differentiation with respect to the argument and
$D_{\mu}$ is the covariant derivative with respect to $g_{\mu\nu}$.
The corresponding junction conditions relating quantities on the
brane are as follows [23]
\begin{equation}
\lim_{\epsilon\rightarrow+0}\Big[K_{\mu\nu}\Big]_{y=-\epsilon}^{y=+\epsilon}=
\frac{f'(R)}{M_{5}^{3}}\Bigg[\tau_{\mu\nu}-\frac{1}{3}g_{\mu\nu}g^{\alpha\beta}\tau_{\alpha\beta}\Bigg]_{y=0}
-\frac{M_{4}^{2}}{M_{5}^{3}}f'(R)\Bigg[R_{\mu\nu}-\frac{1}{6}g_{\mu\nu}g^{\alpha\beta}R_{\alpha\beta}\Bigg]_{y=0}.
\end{equation}
To formulate cosmological dynamics on the brane, we assume the
following line element
\begin{equation}
ds^{2}=-n^{2}(y,t)dt^{2}+a^{2}(y,t)\gamma_{ij}dx^{i}dx^{j}+b^{2}(y,t)dy^{2}
\end{equation}
where $\gamma_{ij}$ is a maximally symmetric 3-dimensional metric
defined as $\gamma_{ij}=\delta_{ij}+k\frac{x_{i}x_{j}}{1-kr^{2}}$
where $k=-1,0,+1$ parameterizes the spatial curvature and
$r^{2}=x_{i}x^{i}$. $y$ is the coordinate of extra dimension and the
brane is located at $y=0$. The junction conditions on the brane now
gives the following expressions
\begin{equation}
\lim_{\epsilon\rightarrow+0}\Big[\partial_{y}n\Big]_{y=-\epsilon}^{y=+\epsilon}(t)
=\frac{2nM_{4}^{2}}{M_{5}^{3}}\Bigg[f'(R)\Big(\frac{\ddot{a}}{n^{2}a}-\frac{\dot{a}^{2}}{2n^{2}a^{2}}
-\frac{\dot{n}\dot{a}}{n^{3}a}-\frac{k}{2a^{2}}\Big)\Bigg]_{y=0}
+\frac{n}{3M_{5}^{3}}\Bigg[f'(R)\Big(2\rho^{(tot)}+3p^{(tot)}\Big)\Bigg]_{y=0}
\end{equation}
and
\begin{equation}
\lim_{\epsilon\rightarrow+0}\Big[\partial_{y}a\Big]_{y=-\epsilon}^{y=+\epsilon}(t)
=\frac{M_{4}^{2}}{M_{5}^{3}}\Bigg[f'(R)\Big(\frac{\dot{a}^{2}}{n^{2}a}+\frac{k}{a}\Big)\Bigg]_{y=0}
-\Bigg[f'(R)\frac{\rho^{(tot)}a}{3M_{5}^{3}}\Bigg]_{y=0}\,.
\end{equation}
If we choose a Gaussian normal coordinate system so that
$b^{2}(y,t)=1$, these equations with non-vanishing components of the
Einstein's tensor in the bulk yield the following generalization of
the Friedmann equation for cosmological dynamics on the brane
$$ \bigg[1+\frac{8}{3}\beta \bigg(H^{2}+\frac{\Phi}{2}+\frac{K}{a^{2}}\bigg)\bigg]^{2}
\bigg(H^{2}-\Phi+\frac{K}{a^{2}}\bigg)=\Bigg\{r\bigg[\bigg(H^{2}+\frac{K}{a^{2}}\bigg)f'(R)\bigg]$$
\begin{equation}
\frac{\kappa_{5}^{2}}{6}\bigg[\rho_{m}+\lambda+\frac{M_{4}^{2}}{2}\bigg(Rf'(R)-f(R)-
6H\dot{R}f''(R)\bigg)\bigg]\Bigg\}^{2}\
,
\end{equation}
where $\rho_{m}$ is the density of ordinary matter on the brane.\,
$\Phi$ is determined by [16]
\begin{equation}
\Phi+2\beta\Phi^{2}=\frac{\Lambda_{5}}{6}+\frac{{\cal{C}}}{a^{4}}\,
\end{equation}
where ${\cal{C}}$ is the bulk black hole mass. In our forthcoming
arguments we set ${\cal{C}}=0$ and by $\Lambda_{5}=0$, we find
$\Phi=0$ and $\Phi=-\frac{1}{2\alpha}$. For simplicity, we choose
$\Phi=0$. Now, if we define the cosmological parameters as\,
$\Omega_{m}=\frac{\kappa^{2}_{4}\rho_{m_{0}}}{3H^{2}_{0}}$,\,
$\Omega_{r_{c}}=\frac{1}{4r_{c}^{2}H^{2}_{0}}$,\,\,
$\Omega_{\alpha}=\frac{8}{3}\alpha H^{2}_{0}$\, and \,
$\Omega_{curv}=\frac{\rho_{curv}\kappa^{2}_{4}}{3H^{2}_{0}}$,\, then
the Friedmann equation on the brane can be expressed as follows
\begin{eqnarray}
E^{2}(z)=\frac{1}{f'(R)}\Big[-2\sqrt{\Omega_{r_{c}}}E(z)\big[1+\Omega_{\alpha}E^{2}
(z)\big]+\Omega_{m}(1+z)^{3}+\Omega_{curv}\Big]
\end{eqnarray}
where $E(z)\equiv\frac{H}{H_{0}}$ and we have set $\lambda=0$.
Evaluating the Friedmann equation at $z=0$ gives a constraint
equation on the cosmological parameters of the model as follows
\begin{eqnarray}
1+2\sqrt{\Omega_{r_{c}}}(1+\Omega_{\alpha})=\Omega_{m}+\Omega_{curv}\,\,.
\end{eqnarray}
We note that a contribution to this equation from $f'(R)|_{z=0}$ has
been normalized to unity. This constraint equation implies that the
subspace with $\Omega_{m}+\Omega_{curv}<1$ is unphysical in the
model parameter space. It is important to note that this model can
be constraint in confrontation with observational data. As we will
show this model mimics the $\Lambda$CDM model by setting
$\Omega_{m}\simeq0.26$,\, $\Omega_{curv}\simeq 0.75$,\,
$\Omega_{r_{c}}\simeq 0.0001$\, and $\Omega_{\alpha}\simeq 0.01$ .

In this framework, the total energy density can be defined as
follows
\begin{equation}
\rho_{tot}=\rho_{m}+\rho_{curv},
\end{equation}
where by definition
\begin{equation}
\rho_{curv}=\frac{M_{4}^{2}}{2}\Big(Rf'(R)-f(R)-6H\dot{R}f''(R)\Big).
\end{equation}
The continuity equation for $\rho_{curv}$ is [24]
\begin{equation}
\dot{\rho}_{curv}+3H(\rho_{curv}+p_{curv})=\frac{3H_{0}^{2}\Omega_{M}\dot{R}f''(R)}{[f'(R)]^{2}}\
a^{-3}\ ,
\end{equation}
where a dot denotes $\frac{d}{d\tau}$ and a prime marks
$\frac{d}{dR}$. Also $\Omega_{M}$ is the present day matter density
parameter. So we can deduce
$$p_{curv}=-\frac{M_{4}^{2}}{6H}\Bigg\{R\dot{R}f''(R)-6\dot{H}\dot{R}f''(R)-6H\ddot{R}f''(R)
-6H\dot{R}^{2}f'''(R)$$
\begin{equation}
+3RHf'(R)-3Hf(R)-18H^{2}\dot{R}f''(R)\Bigg\}+\frac{H_{0}^{2}\Omega_{M}\dot{R}f''(R)}{H[f'(R)]^{2}}a^{-3}.
\end{equation}
Using equations (16), (17) and (18) we find the \emph{curvature
fluid} equation of state parameter as follows
$$\omega_{curv}=-1-\frac{1}{3H}\frac{R\dot{R}f''(R)-6\dot{H}\dot{R}f''(R)
-6H\ddot{R}f''(R)-6H\dot{R}^{2}f'''(R)}{Rf'(R)-f(R)-6H\dot{R}f''(R)
}$$
\begin{equation}
+\frac{2}{M_{4}
H[f'(R)]^{2}}\frac{H_{0}^{2}\Omega_{M}\dot{R}f''(R)a^{-3}}{Rf'(R)-f(R)-6H\dot{R}f''(R)}\,
.
\end{equation}
On the other hand we know that
\begin{equation}
\dot{\rho}_{m}+3H(\rho_{m}+p_{m})=0 ,
\end{equation}
where
\begin{equation}
\rho_{m}=\rho_{0m}(\frac{a_{0}}{a})^{3}=\rho_{0m}(1+z)^{3}.
\end{equation}
So the total equation of state parameter is given as follows
$$\omega_{total}=-1-\frac{1}{3H}\frac{R\dot{R}f''(R)-6\dot{H}\dot{R}f''(R)
-6H\ddot{R}f''(R)-6H\dot{R}^{2}f'''(R)-6\rho_{0}(\frac{a_{0}^{3}}{M_{4}^{2}a^{4}})(\dot{a}-Ha)}
{Rf'(R)-f(R)-6H\dot{R}f''(R)
+\frac{2\rho_{0}}{M_{4}^{2}}(\frac{a_{0}}{a})^{3}}$$
\begin{equation}
+\frac{2}{M_{4}^{2}
H[f'(R)]^{2}}\frac{H_{0}^{2}\Omega_{M}\dot{R}f''(R)a^{-3}}{Rf'(R)-f(R)
-6H\dot{R}f''(R)+\frac{2\rho_{0}}{M_{4}^{2}}(\frac{a}{a_{0}})^{-3}}\,
,
\end{equation}
which includes the matter energy density and pressure too. Albeit,
if we consider a braneworld without ordinary matter content, then
the right hand side of equation (17) vanishes and we find
\begin{equation}
\omega=-1-\frac{1}{3H}\frac{R\dot{R}f''(R)-6\dot{H}\dot{R}f''(R)
-6H\ddot{R}f''(R)-6H\dot{R}^{2}f'''(R)}{Rf'(R)-f(R)-6H\dot{R}f''(R)
}.
\end{equation}
According to recent observational data including the type Ia
supernovae Gold dataset, there exists the possibility that the
effective equation of state parameter evolves from larger than $-1$
(non-phantom phase) to less than -1 (phantom phase, in which
super-acceleration is realized)\,. This means that $w_{eff}$ crosses
$-1$ line (the phantom divide line) smoothly. A number of attempts
to realize the crossing of the phantom divide have been made in the
framework of general relativity. For instance, we could mention
scalar-tensor theories with the nonminimal gravitational coupling
between a scalar field and the scalar curvature or that between a
scalar field and the Gauss-Bonnet term, one scalar field model with
nonlinear kinetic terms or a non-linear higher-derivative one,
phantom coupled to dark matter with an appropriate coupling, the
thermodynamical inhomogeneous dark energy model, multiple kinetic
k-essence, multi-field models (two scalar fields model, "quintom"
consisting of phantom and canonical scalar fields), and the
description of those models through the Parameterized Post-Friedmann
approach, or a classical Dirac field or string-inspired models,
non-local gravity, a model in loop quantum cosmology and a general
consideration of the crossing of the phantom divide ( see for
instance [25] and references therein) and very recently crossing
with Lorentz invariance violating fields [26] . In addition, there
are interesting models of modified gravity realizing the crossing of
the phantom divide too [27]. In the present paper, we study
possibility of realization of the phantom-like behavior and phantom
divide line crossing in our modified $GBIG$ model without
introducing any phantom matter on the brane or in the bulk. For this
purpose, we adopt some reliable ansatz for $f(R)$ and $a(t)$ as
follows.

\subsection{Phantom-Like behavior and crossing of the phantom
divide} By phantom-like behavior one means an effective energy
density which is positive and grows with time and its equation of
state parameter stays always less than $-1$. In this subsection we
study possible realization of the phantom-like behavior in this
UV/IR-complete theory. To do this end, we define effective energy
density and effective equation of state parameter as basic
ingredients of our analysis. To find the effective energy density we
can use
\begin{equation}
H^2=\frac{\kappa_{4}^{2}}{3}(\rho_{m}+\rho_{eff})
\end{equation}
and then we rewrite our Friedmann equation (11) in this way to find
\begin{equation}
\rho_{eff}=-\frac{3}{r\kappa_{4}^{2}f'(R)}\Big[(1+\frac{8}{3}\alpha
H^2)H-\frac{\kappa_{5}^{2}}{6}(\rho_{m}+\rho_{curve})\Big]-\rho_{m}.
\end{equation}
Using the continuity equation
\begin{equation}
\dot{\rho}_{eff}+3H(1+\omega_{eff})\rho_{eff}=0\,,
\end{equation}
since
\begin{tiny}
$$
p_{eff}=\frac{-1}{rH\kappa_{4}^{2}f'(R)}\Bigg\{(1+8\alpha
H^{2})\dot{H}-\frac{\kappa_{5}^{2}}{6}(\dot{\rho}_{m}
+\dot{\rho}_{curv})-\frac{r\kappa_{4}^{2}}{3}f'(R)\dot{\rho}_{m}-\frac{f''(R)}{f'(R)}\dot{R}\Big[(1+\frac{8}{3}\alpha
H^2)H-\frac{\kappa_{5}^{2}}{6}(\rho_{m}+\rho_{curv})\Big]\Bigg\}
$$
\begin{equation}
+\frac{3}{r\kappa_{4}^{2}f'(R)}\Big[(1+\frac{8}{3}\alpha
H^2)H-\frac{\kappa_{5}^{2}}{6}(\rho_{m}+\rho_{curv})\Big]+\rho_{m}\,,
\end{equation}
\end{tiny}
we find
\begin{tiny}
\begin{equation}
1+\omega_{eff}=\frac{1}{3H}\Bigg\{\frac{(1+8\alpha
H^{2})\dot{H}-\frac{\kappa_{5}^{2}}{6}(\dot{\rho}_{m}
+\dot{\rho}_{curv})-\frac{r\kappa_{4}^{2}}{3}f'(R)\dot{\rho}_{m}-\frac{f''(R)}{f'(R)}\dot{R}\big[(1+\frac{8}{3}\alpha
H^2)H-\frac{\kappa_{5}^{2}}{6}(\rho_{m}+\rho_{curv})\big]}{(1+\frac{8}{3}\alpha
H^2)H-\frac{\kappa_{5}^{2}}{6}(\rho_{m}+\rho_{curv})+\frac{r\kappa_{4}^{2}}{3}f'(R)\rho_{m}}\Bigg\}
\end{equation}
\end{tiny}
Now to proceed further, we should specify the form of $f(R)$.
Extended theories of gravity based on four dimensional $f(R)$
scenarios should follow closely the expansion of a $\Lambda$CDM
universe [7,28,29] and also could have distinctive signatures on the
large scale structure of the universe [30,31]. During the last few
years, several methods have been proposed to reconstruct the shape
of $f(R)$ from observations [32,24]. This has been done, for
example, by using the dependence of the Hubble parameter with
redshift which can be retrieved from astrophysical observations
[24]. Among these attempts, the models presented in Ref. [28] are
used in our forthcoming arguments to study phantom-like behavior and
crossing of the phantom divide line in our modified GBIG model. We
proceed as follows: we start with the following observationally
suitable ansatz for $f(R)$ [28]
\begin{equation}
f(R)=R-R_{c}\frac{\alpha(R/R_{c})^{n}}{1+(R/R_{c})^{n}},
\end{equation}
where
\begin{equation}
R=-6(\dot{H}+2H^{2})
\end{equation}
for an spatially flat FRW type universe. Both  $\alpha$ and $R_{c}$
are free positive parameters. We also adopt the following
observationally reliable ansatz [33]
\begin{equation}
a=\bigg(t^2+\frac{t_{0}}{1-\nu}\bigg)^{\frac{1}{1-\nu}}
\end{equation}
and we translate all of our cosmological dynamics equations in terms
of red-shift using the following relation between $a$ and $z$
\begin{equation}
1+z=\frac{a_{0}}{a}.
\end{equation}
Figure $1$ shows the behavior of the effective energy density,
$\rho_{eff}$, versus redshift. The effective energy density grows
with time and $\rho_{eff}>0$ always. These are necessary conditions
for phantom-like behavior but not sufficient: the effective equation
of state parameter should stay below $-1$. Figure $2$ shows the
behavior of $1+\omega_{eff}$ versus redshift. The universe enters
the phantom phase in the near past and currently it is in the
phantom phase. The transition to the phantom phase has occurred  at
$z\simeq 0.25$. We note that based on some observations, crossing of
the phantom divide line by the equation of state parameter occurs at
$z\simeq 0.25$. For instance the Gold sample mildly favors a
crossing of the phantom divide line at $z\simeq0.25$ while no such
trend appears for the SNLS data ( see for instance the paper by
Nesseris and Perivolaropoulos in Ref. [25]).
\begin{figure}[htp]
\begin{center}\includegraphics{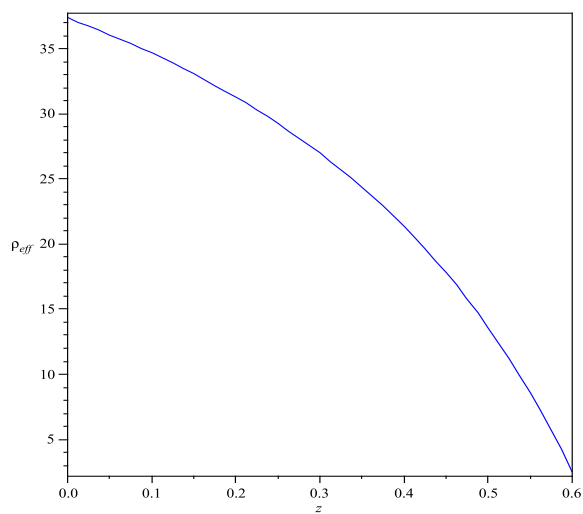} \vspace{6cm}
\end{center}
 \caption{\small { Variation of the effective energy density versus the redshift. }}
\end{figure}

\begin{figure}[htp]
\begin{center}\includegraphics{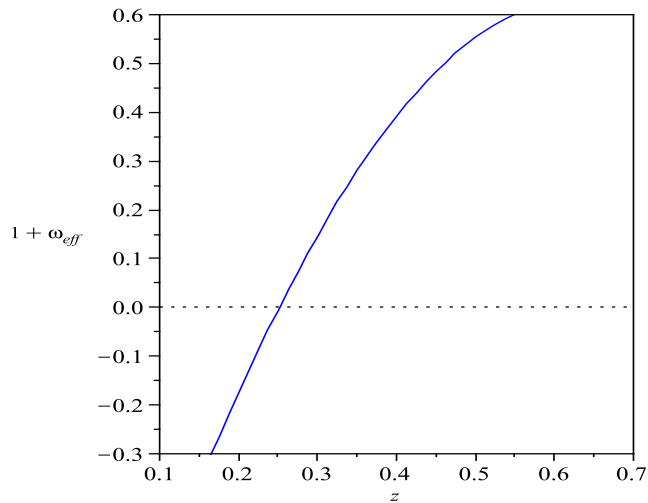} \vspace{6.2cm}
\end{center}
 \caption{\small { Variation of $1+\omega_{eff}$ versus the redshift. }}
\end{figure}
Up to now we have shown that this modified GBIG scenario realizes
the phantom-like behavior without introducing any phantom matter on
the brane or in the bulk. This phantom-like behavior is a mimicry of
$\Lambda$CDM scenario [6,7]. Now, the deceleration parameter defined
as
\begin{equation}
q=-[\frac{\dot{H}}{H^{2}}+1]
\end{equation}
takes the following form in our model
\begin{equation}
q=-\Big[1-\frac{r\dot{R}f''(R)-\frac{\kappa_{5}^{2}}{6H^2}(\dot{\rho}_{m}
+\dot{\rho}_{curv})}{1+8\alpha H^2+2rHf'(R)}\Big]\,.
\end{equation}
Figure $3$ shows behavior of $q$ versus $z$. The universe has
entered an accelerating phase in the past at $z\simeq0.84$.
\begin{figure}[htp]
\begin{center}\includegraphics{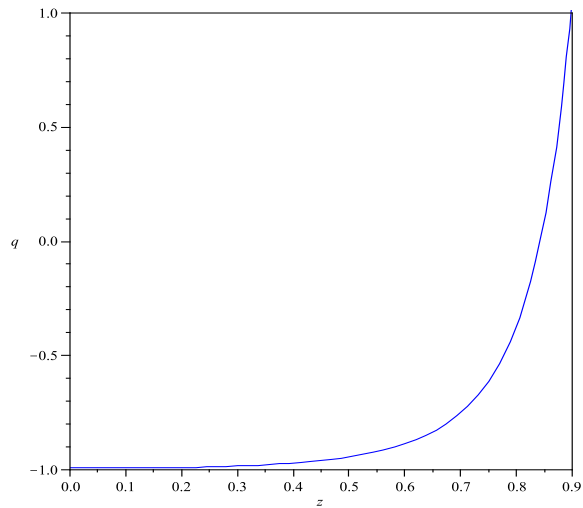} \vspace{7cm}
\end{center}
 \caption{\small { Variation of the deceleration parameter versus redshift. }}
\end{figure}
The variation of $H$ with redshift gives another part of important
information about the cosmology of this model. It is given by
\begin{equation}
\frac{\dot{H}}{H_{0}^{2}}=-\frac{1}{H_{0}^2}\frac{r\dot{R}H^{2}f''(R)
-\frac{\kappa_{5}^{2}}{6}(\dot{\rho}_{m}+\dot{\rho}_{curv})}{1+8\alpha
H^2+2rHf'(R)}.
\end{equation}
Figure $4$ shows the variation of $\frac{\dot{H}}{H_{0}^{2}}$ versus
redshift. Since $\dot{H}<0$ always, the model universe described
here will not acquire super-acceleration and big-rip singularity in
the future.
\begin{figure}[htp]
\begin{center}\includegraphics{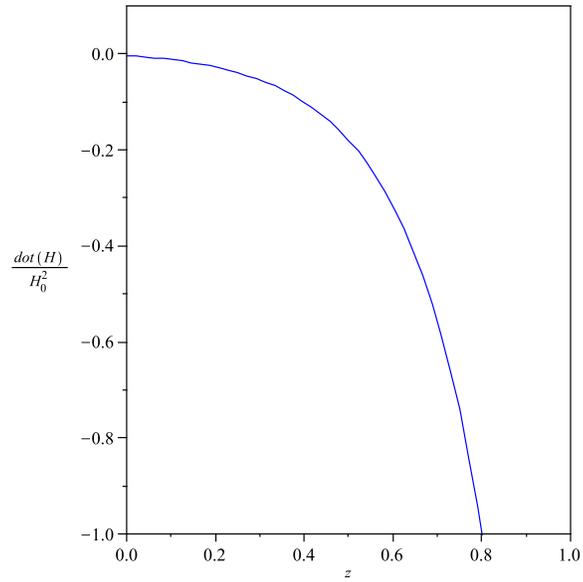} \vspace{6cm}
\end{center}
 \caption{\small { Variation of $\frac{\dot{H}}{H_{0}^{2}}$ versus the redshift. }}
\end{figure}
Since we have not introduced any phantom field on the brane or in
the bulk, we expect that the null energy condition to be respected
at least in some subspaces of the model parameter space. Figure $5$
shows the status of the null energy condition in this setup. As this
figure shows, phantom-like behavior occurs in this setup without
violation of the null energy condition. We note that this model
mimics a $\Lambda$CDM model in several respects and in this sense it
is important to confront its parameter space with recent observation
to see its viability.

\begin{figure}[htp]
\begin{center}\includegraphics{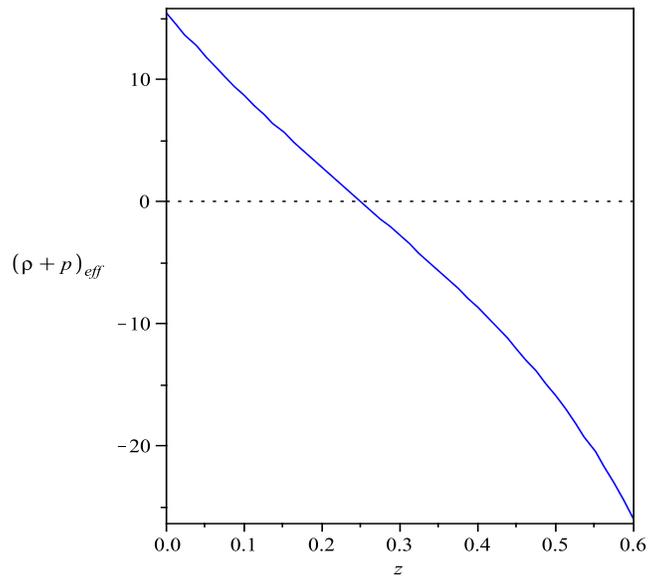} \vspace{6cm}
\end{center}
 \caption{\small { Variation of $\rho_{eff}+P_{eff}$ versus the redshift.
 Null energy condition is respected in the phantom phase of the model. }}
\end{figure}
Finally we focus on the appropriate ranges of $n$ that crossing of
the phantom divide line can be realized.  Our analysis shows that
$\omega_{curv}$ as defined by (19) crosses the phantom divide line
smoothly if \,$2.42\leq n \leq4.68$. Also, $\omega_{total}$ crosses
the phantom divide line if $2.44\leq n \leq 4.74$. And finally,
$\omega$ as defined in equation (23), crosses the phantom line if
$2.43 \leq n\leq 5.15$. In our model, this crossing occurs at
$z\simeq 0.27$ which lies well in the vicinity of the
observationally supported value of $z\simeq 0.25$. Figure $6$ shows
the behavior of these equation of state parameters versus $z$ for
$n=4$. As this figure shows, crossing of the phantom divide evolves
from quintessence to phantom phase in the same way as observations
suggest.
\begin{figure}[htp]
\begin{center}\includegraphics{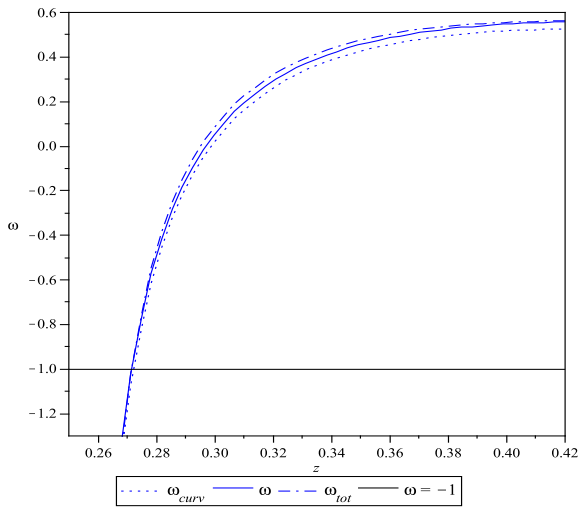} \vspace{6cm}
\end{center}
 \caption{\small { Smooth crossing of the phantom divide line. }}
\end{figure}
\section{Summary}
In this paper we have constructed a modified GBIG scenario: it
contains Gauss-Bonnet term as the UV sector of the theory and
Induced Gravity effect as IR side of the model. The induced gravity
on the brane is modified in the spirit of $f(R)$ gravity. The
cosmological dynamics in this braneworld setup mimics a $\Lambda$CDM
model in several respects: it realizes phantom-like behavior without
introducing any phantom matter on the brane. The effective energy
density increases with cosmic time and effective equation of state
parameter crosses the phantom divide line smoothly in the same way
as observations suggest. This crossing behavior occurs at $z\simeq
0.27$ which lies well in the vicinity of the observationally
supported value of $z\simeq 0.25$. The phantom-like behavior
realized in this model happens without violation of the null energy
condition in the phantom phase. Finally, to have smooth crossing of
the phantom divide line by $\omega_{eff}$, the value of parameter
$n$ in the Hu-Sawicki model should lie in the range $2.42\leq n
\leq4.68$.

\end{document}